\documentclass[letterpaper,10pt]{article}  
\usepackage{opex3}
\usepackage[latin1]{inputenc}
\usepackage[dvips]{hyperref}
\usepackage{amstext}
\begin{document}

\title{Imaging a vibrating object by Sideband Digital Holography}

\author{F. Joud$^1$, F. Lalo\"{e}$^1$,  M. Atlan$^{2}$, J. Hare$^1$ and M. Gross$^1$.}

\address{ $^1$~Laboratoire Kastler Brossel -- UMR 8552
\'Ecole Normale Sup\'erieure, UPMC, CNRS\\
 24 rue Lhomond ; 75231 Paris Cedex 05 ; France\\
 $^2$~D\'epartement de Biologie Cellulaire --- Institut Jacques
Monod, UMR 7592, CNRS, Univ. Paris 6 and 7 ---
2 Place Jussieu ; 75251 Paris Cedex 05;  France}

\email{gross@lkb.ens.fr}

\begin{abstract}
We obtain quantitative measurements of the oscillation amplitude
of vibrating objects by using sideband digital holography. The
frequency sidebands on the light scattered by the object, shifted
by $n$ times the vibration frequency, are selectively detected by
heterodyne holography, and images of the object are calculated for
different orders $n$. Orders up to $n=120$ have been observed,
allowing the measurement of amplitudes of oscillation that are
significantly larger than the optical wavelength. Using the
positions of the zeros  of intensity for each value of $n$, we
reconstruct the shape of vibration the object.
\end{abstract}




%
%
%
Holographic imaging is based  on interferences between a signal
optical field and a reference beam; it provides an accurate method
to image the vibration of various objects. Powell and Stetson
\cite{powell1965iva} have shown that the {\sl time-averaged}
hologram of an harmonically vibrating object involves the Bessel
function $J_{0}(z)$, where $z$ is the phase modulation amplitude.
In the backscattering geometry, $z=4 \pi A/\lambda$ where $A$ is
the mechanical  amplitude of vibration and $\lambda$ the optical
wavelength. Image reconstruction then provides a direct mapping of
the amplitude $A$ at various points of the object, in the form of
dark fringes appearing around points where $J_{0}(z)$ is zero.

Picard et al. \cite{picart2003tad} have simplified the processing
of the data by performing Time Averaged Digital Holography (TADH)
with a CCD camera, leading to numerous recent developments
\cite{zhang2004vap,asundi2006tal,singh2007dcm,demoli2004dhs,demoli2006rtm}.
Nevertheless, the quantitative measurement of vibration amplitudes
remains limited to situations where fringes can be counted, which
implies that the amplitudes of vibration must be relatively small;
otherwise, it becomes difficult to count many narrow fringes, and
even impossible when they are smaller than the optical resolution.

\begin{figure}[b]
\centering
\includegraphics[width=12cm]{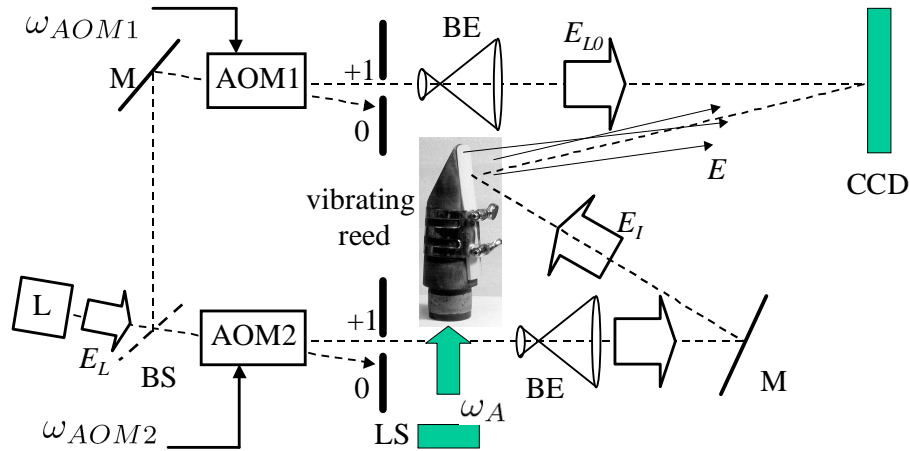}
\caption{Setup. L: main laser; AOM1, AOM2: acousto-optic
modulators; M: mirror; BS: beam splitter; BE: beam expander; CCD:
CCD camera; LS: loud-speaker exiting the clarinet reed at
frequency $\omega_A/2\pi$. } \label{setup}
\end{figure}

%

In this letter, we show how Sideband Digital Holography (SDH) can
overcome the problem. Sideband holography was already demonstrated
in 1971 by C.C.~Aleksoff, in a pioneering experiment
\cite{aleksoff1971} where the reference beam was frequency-shifted
by an ultrasonic diffraction cell; selecting a given frequency
resulted in a selection of a given sideband of the signal
backscattered by the object. The experiment was done with
holographic plates. Here,  we use digital heterodyne holography
\cite{Leclerc2000}; both reference and illumination laser beams
are frequency shifted, and the selection of one sideband is
obtained by a proper detuning of one of the beams followed by
heterodyne detection on a CCD camera; this technique provides much
more flexibility and leads to accurate quantitative measurements.
As a demonstration, inspired by the work described in
\cite{pinard2003mqa, boutillon2003} we perform an experiment where
the vibrating object is the reed of a clarinet; in addition to the
intrinsic interest of such an object, with its possible musical
implications, a reed provides a test system that is particularly
well adapted to our purposes, with typical vibration amplitudes of
the order of 0.1~mm. We will see that, even with smaller
amplitudes, there is no difficulty in obtaining holographic images
corresponding to the $n^{\text{th}}$ sideband with $n$ up to $100$
or more.

The experimental setup is schematically shown in Fig.~\ref{setup}.
A laser beam, with  wavelength $\lambda\simeq 650\;\text{nm}$
(angular frequency $\omega_L$) is split into a local oscillator
beam ($E_{LO}$) and an illumination beam ($E_{I}$); their angular
frequencies $\omega_{LO}$ and $\omega_{I}$ are tuned by using two
acousto-optic modulators (Bragg cells with a selection of the
first order  diffraction beam) AOM1 and AOM2:
$\omega_{LO}=\omega_{L} + \omega_\text{AOM1}$ and
$\omega_{I}=\omega_{L} + \omega_\text{AOM2}$ where
$\omega_\text{AOM1,2}/2\pi \simeq 80\;\text{MHz}$. The clarinet
reed is attached to a clarinet mouthpiece and its vibration  is
driven by a sound wave propagating inside the mouthpiece, as in
playing conditions, but in our experiment the sound wave is
created by a loudspeaker excited at frequency $\omega_{A}$ and has
a lower intensity than inside a clarinet; no attempt has been made
to reproduce the mechanical effect of the lip of the player. The
excitation frequency is adjusted to be resonant with the first
flexion mode (2143~Hz) of the reed. The phase of the field $E$
backscattered by any element of the reed is then phase modulated
at frequency $\omega_{A}$, so that $E$ can be expanded into
carrier ($E_0$) and sideband components ($E_n$) as:
\begin{eqnarray}\label{Eq_1_}
  E(t)={\cal E} e^{j \omega_I t} e^{j z \sin(\omega_A t)} =
\sum_{n=-\infty}^{\infty} E_n(t) \\ 
 E_n(t)= {\cal E} J_{n}(z) e^{j \omega_{L}t} e^{j (\omega_{AOM2}+n \omega_{A} )t}
\end{eqnarray}
where ${\cal E}$ is  the complex field of the object in the
absence of modulation, $z=4 \pi A/\lambda$ is the phase modulation
amplitude, $A$  the amplitude of vibration of the particular
element, and $J_n$ is the $n$-th order Bessel function of the
first kind (with $J_{-n}(z)=(-1)^{n}\; J_{n}(z)$ for  integer $n$
and  real $z$).
\begin{figure}[bt]
\centering
\includegraphics[width=12cm]{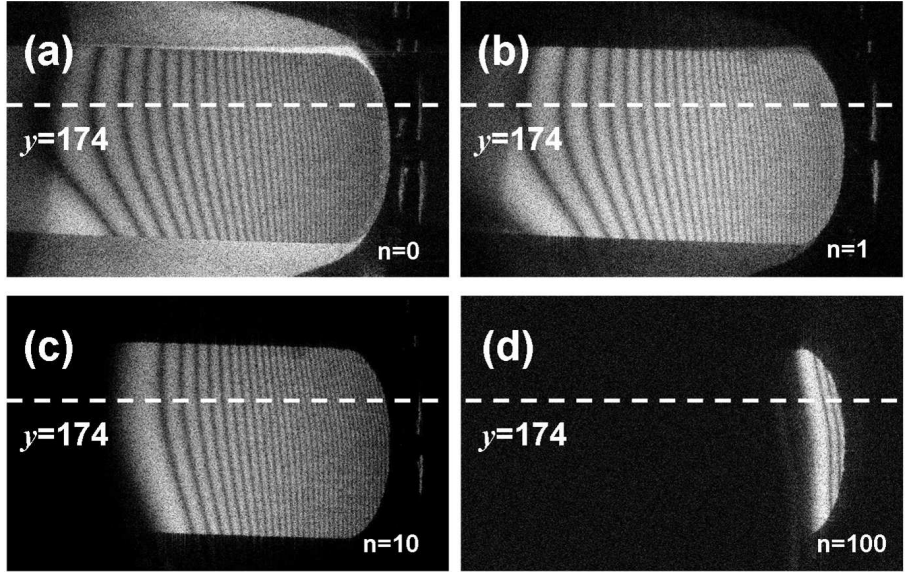}
\caption{Reconstructed holographic  images of a clarinet reed
vibrating at frequency $\omega_{A}/2\pi=2143\ \text{Hz}$
perpendicularly to the plane of the figure. Figure~(a) shows the
carrier image obtained for $n=0$; Fig.~(b)-(d) show the frequency
sideband  images respectively for $n=1$, $n=10$, and $n=100$. A
logarithmic grey scale has been used.}
\label{manip6}
\end{figure}

To selectively  detect the $n^\text{th}$ sideband we use 4
phase shifting holography \cite{Yamaguchi1997} and adjust
the frequency $\omega_\text{AOM1}$ to fulfil the condition:
\begin{equation}\label{Eq_omega_}
\omega_\text{AOM2}-\omega_\text{AOM1}- n \;\omega_{A}=\omega_\text{\sc ccd}/4
\end{equation}
where $\omega_\text{\sc ccd}$ is the CCD camera frame frequency
and $n$ is a (positive or negative) integer; the phase of the
modulated signal is then shifted by $90^\circ$  from one CCD image to
the next \cite{atlan2007aps}. The sideband complex hologram signal
$H_n$ provided by each pixel of the camera, proportional to the
local sideband  complex field, is  obtained by 4-phase
demodulation:
\begin{equation}\label{Eq_omega_2}
 H_n= (I_{0}-I_{2})+ j(I_{1}-I_{3}) \ ,
\end{equation}
where $ I_{0},\cdots I_{3}$ are 4 consecutive intensity images
digitally  recorded by the CCD camera (Pike: Allied vision
Technology Inc, 12 bits, $\omega_\text{\sc ccd}/2\pi =
10\;\text{Hz}$, exposure time $100\;\text{ms}$, $1340\times 1024$
pixels). The LO beam is slightly angularly shifted ($\sim 1^\circ$)
with respect to the beam originating from the reed; in this way,
the zero order and twin-image aliases \cite{Cuche00} can be
suppressed and an ultimate detection sensitivity can be reached
\cite{gross_07,gross_08}.

%
%

\begin{figure}[tbp]
\centering
\includegraphics[width=13cm]{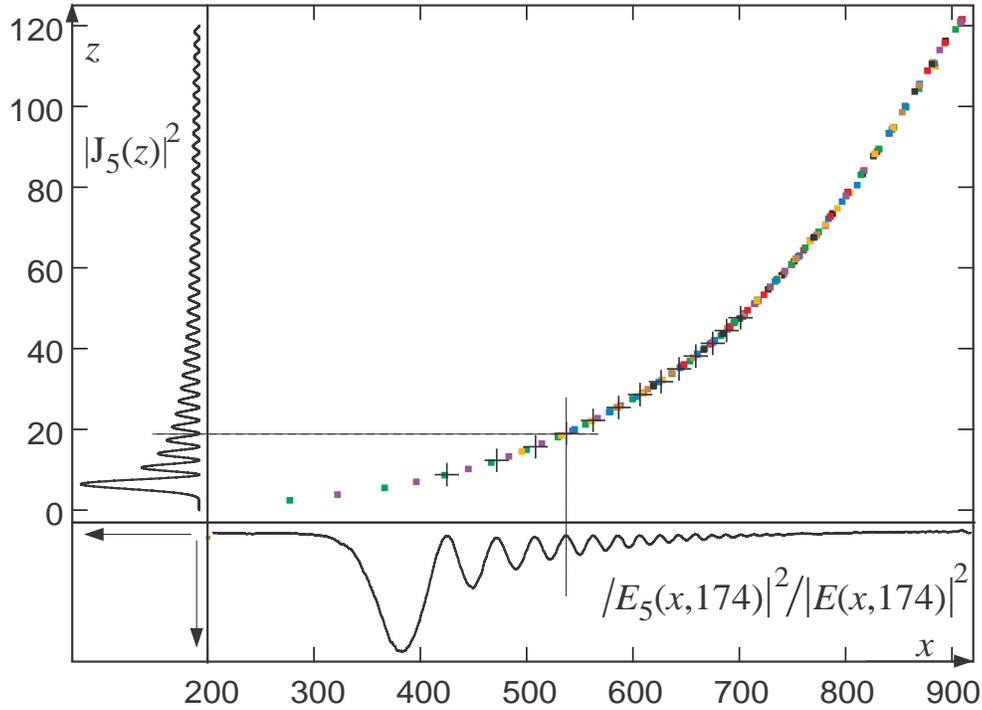}
\caption{ A slice of the data along the $y=174$ line is used in
this figure; the $x$ horizontal  axis gives the pixel index (100
pixels correspond to 3.68~mm.), the vertical axis the phase
modulation amplitude $z$. The lower part of the figure shows the
normalized signal $|E_n(x)|^2/|E(x)|^2$ for a particular sideband
order $n=5$, with a downwards axis; the left part shows the
corresponding square of the Bessel function $|J_5(z)|^2$ with a
leftward axis. The zeroes of the two curves are put in
correspondence, which provides the points in the central figure; a
different color is used for each harmonic order
$n=0,1,5,10...100$; the crosses correspond to $n=5$. The
juxtaposition of the points for all values of $n$ gives an
accurate  representation of the amplitude of vibration $A$ as a
function of $x$.} \label{fig_zeros}
\end{figure}

From the complex holograms, the images of the reed are
reconstructed by the standard convolution method
\cite{Schnars_Juptner_94}. To simplify the Fourier transform
calculation, the $1340\times 1024$ pixels data are truncated to
$1024\times 1024$. Because of the angular shift, the image of the
reed lies in the upper side of the reconstruction grid. In figure
\ref{manip6}, for pixels ranging from $200$ to $1023$ along the
$x$ axis and $0$ to $511$ on the $y$ axis, we show the
reconstructed images for different sideband orders $n$; only the
square modulus of the  complex signal is retained (the phase
coming from $\cal{ E}$ is ignored), so that we obtain intensity
images, proportional to $|E_n|^2$ for each different sideband.

Figure ~\ref{manip6}(a) is obtained at the unshifted carrier
frequency ($n=0$). The left side of the reed is attached to the
mouthpiece, and the amplitude of vibration is larger at the tip of
the reed on the right side; in this region the fringes become
closer and closer and difficult to count. The mouthpiece is also
visible, but without fringes since it does not vibrate. Similar
images of clarinet reeds have been obtained in
\cite{pinard2003mqa,boutillon2003}, with more conventional
techniques and lower image quality. Figures
~\ref{manip6}(b)-\ref{manip6}(d) show images obtained for three
sideband frequencies. As expected, the non-vibrating mouthpiece is
no longer visible. Figure ~\ref{manip6}(b) shows the $n=1$
sideband image, with $J_1$ fringes that are slightly shifted with
respect to those of $J_0$. Figure \ref{manip6}(c) shows the order
$n=10$ and (d) the order $n=100$ (e).  The left side region of the
image remains dark because, in that region, the vibration
amplitude is not sufficient to generate these sidebands,
$J_{n}(z)$ being evanescent for $z<n$.

We have performed cuts of the reconstructed images signal
$|E_n(x,y|^2$ obtained for different sideband orders $n$ along the
horizontal line $y=174$ (this value has been chosen because it
corresponds to a region where the fringes are orthogonal to the
$y$ axis). To reduce the effect of speckle fluctuations, the
intensity signal $|E_n(x,y=174)|^2$ has been averaged over 20
pixels in the $y$ direction ($y=164\ldots 184$). Moreover, since
the illumination of the reed is not uniform, we have normalized
the sideband signal by the reconstructed image intensity $|E|^2$
obtained at the carrier frequency ($n=0$) without mechanical
excitation of the reed (loud-speaker off). One example of the
normalized signals $|E_n|^2/|E|^2 $ along the cut is shown for
$n=5$, in the lower part of Fig.~\ref{fig_zeros}, with a downwards
intensity axis. In this case, and for many other values of $n$, we
have checked that the normalized curves vary as the square of
Bessel functions, with a first fringe that moves to higher values
when $n$ increases (it occurs when $z\simeq n$ and the amplitude
of vibrations increases with $x$). For instance, for $n=0$, which
corresponds to  standard TADH, the variations of $|J_0(z)|^2$ show
a large number of nodes and anti-nodes, with decreasing visibility
for increasing $x$; above $x=700$, the fringes are no longer
visible. For $n=20$, the first fringe, which corresponds to
$z\simeq n=20$, is located near $x\simeq560$. Going to $n=60$
pushes the first fringe to $x\simeq 755$, where the counting of
fringes with TADH would no longer be possible. In a more general
way, the first fringe on the $n^\text{th}$ sideband image provides
a convenient marker for the region $z\simeq n$.

For a more precise treatment of the data, we use  the position of
the antinodes, which is insensitive to inhomogeneities of
illumination; no normalization is then required. To build the
central part of Fig.~\ref{fig_zeros}, a similar linear cut at
$y=174$ is processed, for the values $n = 0,1,5,10,\ldots 100$.
The $x$ locations of the successive minima of the signal
$|E_n(x)|^2$ are plotted against the  zeros of the corresponding
$|J_n(z)|^2$.  To illustrate the way the points are  obtained, we
show in the left part of the Fig.~\ref{fig_zeros} the expected
signal $|J_5(z)|^2$, to be compared to the signal in the lower
part. For all observed sideband order $n$ ranging from 0 to 100,
all the points displayed on Fig.~\ref{fig_zeros} fall on a well
defined curve, which provides the vibration amplitude $z$ as
function of the location $x$.  The excellent consistency of the
overlapping data sets demonstrates the validity of expansions
Eq.~\ref{Eq_1_} for any order $n$. Therefore the figure provides a
direct and accurate visualization of the map of the maximal
elongations of the reed, from the left part clamped on the
mouthpiece to the tip on the right; here, the maximum amplitude is
$z\simeq 120$ radians, corresponding to $A\simeq
6.2\;\mu\text{m}$.

To conclude, recording separately different  sideband Fourier
components of the light backscattered by a vibrating object gives
access to a large amount of information; this can be used to
remove ambiguities and inaccuracies in measurements of large
modulation amplitude. Using large sideband  order $n$ actually
makes the calibration of the vibration amplitude $z$
straightforward: the first fringe of a high order image can be
used as a marker. This marker is easy to locate since it
corresponds to a fast transition from zero signal, and moreover is
brighter and  broader. Performing sideband holography with many
orders $n$ gives a direct and accurate access to the shape of the
vibration amplitude and its position dependence. Figure
\ref{fig_zeros} shows only a representation of a slice of the data
along one given axis, but more information is available if one
changes the value of $y$ or chooses another direction, for
instance parallel to $y$ to accurately map the transverse
variations of the amplitude $A$. Sideband digital holography
therefore opens up a variety of new possibilities in holography.

The authors acknowledge the support of French National Research
Agency (ANR-05-NANO-031) and the ``Centre de compétence
NanoSciences Île de France'' (C'nano IdF).

\end{document}